# Massive MIMO with Imperfect Channel Covariance Information

Emil Björnson*, Luca Sanguinetti†‡, Merouane Debbah‡§

*Department of Electrical Engineering (ISY), Linköping University, Linköping, Sweden
†Dipartimento di Ingegneria dell'Informazione, University of Pisa, Pisa, Italy
‡Large Networks and System Group (LANEAS), CentraleSupélec, Université Paris-Saclay, Gif-sur-Yvette, France
§Mathematical and Algorithmic Sciences Lab, Huawei Technologies, France.

*Abstract*—This work investigates the impact of imperfect statistical information in the uplink of massive MIMO systems. In particular, we first show why covariance information is needed and then propose two schemes for covariance matrix estimation. A lower bound on the spectral efficiency (SE) of any combining scheme is derived, under imperfect covariance knowledge, and a closed-form expression is computed for maximum-ratio combining. We show that having covariance information is not critical, but that it is relatively easy to acquire it and to achieve SE close to the ideal case of having perfect statistical information.

## I. INTRODUCTION

Massive MIMO (multiple input multiple output) is considered a key technology for next generation cellular networks [1]–[3]. The technology evolves conventional base stations (BSs) by using arrays with hundreds of electronically steerable antennas, which enable spatial multiplexing of tens of user equipments (UEs) per cell—in both uplink (UL) and downlink (DL). To this end, the BS needs to learn the channel vector of each of the UEs that it serves. Due to channel fading, each channel vector must be re-estimated frequently over both time and frequency domains.

Practical channels are spatially correlated [4], which implies that the elements of a channel vector are correlated. The covariance matrix of the channel vector characterizes the spatial channel characteristics and is needed to apply an optimal minimum mean squared error (MMSE) channel estimator [2]. Covariance matrix information is also important for resource allocation [5] and to suppress pilot contamination [6], [7]. However, the covariance matrices are commonly assumed to be perfectly known in the massive MIMO literature, which is questionable since the matrix dimensions grow with the number of antennas and the statistics change over time due to mobility. Practical covariance estimates are imperfect since the number of observations may be at the same order as the number of antennas. One promising method to estimate a large-dimensional covariance matrix, using such a small number of observations, is to regularize the sample covariance matrix [8], [9]. In this paper, we utilize this method and investigate two different approaches to estimate the various covariance matrices needed for MMSE estimation in multi-cell systems. We derive mean-squared error (MSE) and SE expressions and evaluate the two approaches numerically.

This research has been supported by the ERC Starting Grant 305123 MORE, ELLIIT, CENIIT, and by the research project 5GIOTTO funded by the University of Pisa.

## II. SYSTEM MODEL

We consider the uplink of a massive MIMO system with $L$ cells, each comprising a BS with $M$ antennas and $K$ single-antenna UEs. We assume block flat-fading channels where $B_c$ (in Hz) is the coherence bandwidth and $T_c$ (in seconds) is the coherence time. Hence, the channels are static within time-frequency coherence blocks of $\tau_c = B_c T_c$ channel uses. Let $\mathbf{h}_{jlk} \in \mathbb{C}^M$ be the channel from BS $j$ to UE $k$ in cell $l$ within a block. We consider correlated Rayleigh fading channels with

$$\mathbf{h}_{jlk} \sim \mathcal{N}_{\mathbb{C}}\left(\mathbf{0}, \mathbf{R}_{jlk}\right) \qquad (1)$$

where $\mathbf{R}_{jlk} \in \mathbb{C}^{M \times M}$ is the channel covariance matrix. The matrix $\mathbf{R}_{jlk}$ describes the macroscopic effects, including spatial channel correlation and average pathloss in different spatial directions. Independent Rayleigh fading with $\mathbf{R}_{jlk} = \beta_{jlk} \mathbf{I}_M$ is a special case of (1) that is often considered in the literature, but we stress that practical covariance matrices are typically non-diagonal [4]. In practice, the matrices $\{\mathbf{R}_{jlk}\}$ maintain constant over the transmission bandwidth and change slowly in time compared to $\{\mathbf{h}_{jlk}\}$. The measurements in [10] suggest roughly two orders of magnitude slower variations. We assume that the matrices $\{\mathbf{R}_{jlk}\}$ maintain constant for $\tau_s$ channel blocks, where $\tau_s$ can be at the order of thousands.

### A. Channel Estimation: Perfect Covariance Information

We assume that the BSs and UEs are perfectly synchronized and operate according to a pilot-based protocol. Each coherence block contains $\tau_c$ channel uses, whereof $K$ are dedicated for pilots. There are $K$ orthogonal unit-norm pilot sequences, which are reused across the cells. The pilot associated with UE $k$ in cell $j$ is denoted by $\boldsymbol{\phi}_{jk} \in \mathbb{C}^K$ and satisfies $\|\boldsymbol{\phi}_{jk}\|^2 = 1$. The $k$th UE in each cell uses the same pilot. Using a (normalized) total pilot power of $\rho^{\text{tr}}$ per UE, the MMSE estimate of $\mathbf{h}_{jjk}$ takes the following form [11]:

$$\hat{\mathbf{h}}_{jjk} = \mathbf{R}_{jjk} \mathbf{Q}_{jk}^{-1} \mathbf{y}_{jk}^p \qquad (2)$$

with

$$\mathbf{y}_{jk}^p = \mathbf{h}_{jjk} + \sum_{l=1, l \neq j}^{L} \mathbf{h}_{jlk} + \frac{1}{\sqrt{\rho^{\text{tr}}}} \mathbf{N}_j^p \boldsymbol{\phi}_{jk}^\star \qquad (3)$$

and $\mathbf{Q}_{jk} = \mathbb{E}\{\mathbf{y}_{jk}^p (\mathbf{y}_{jk}^p)^{\text{H}}\}$ given by

$$\mathbf{Q}_{jk} = \sum_{l=1}^{L} \mathbf{R}_{jlk} + \frac{1}{\rho^{\text{tr}}} \mathbf{I}_M. \qquad (4)$$

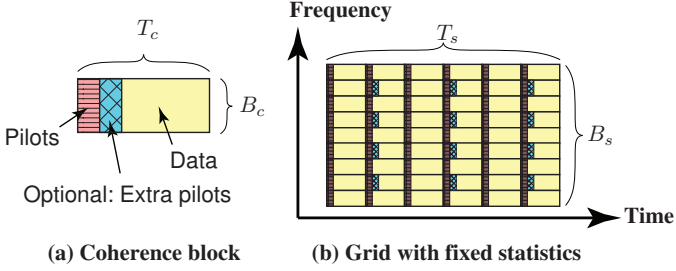

Fig. 1. (a) Some coherence blocks contain extra pilots for covariance matrix estimation; (b) Time-frequency grid over which the channel statistics are fixed.

The estimate $\hat{\mathbf{h}}_{jjk}$ and estimation error $\tilde{\mathbf{h}}_{jjk} = \mathbf{h}_{jjk} - \hat{\mathbf{h}}_{jjk}$ are independently distributed as $\hat{\mathbf{h}}_{jjk} \sim \mathcal{N}_{\mathbb{C}}(\mathbf{0}, \boldsymbol{\Phi}_{jjk})$ and $\tilde{\mathbf{h}}_{jjk} \sim \mathcal{N}_{\mathbb{C}}(\mathbf{0}, \mathbf{R}_{jjk} - \boldsymbol{\Phi}_{jjk})$ with $\boldsymbol{\Phi}_{jjk} = \mathbf{R}_{jjk}\mathbf{Q}_{jk}^{-1}\mathbf{R}_{jjk}$. Note that the MMSE estimator utilizes the channel statistics; that is, the Gaussian distribution, mean value, and covariance matrix. In particular, the BS can only compute the estimate $\hat{\mathbf{h}}_{jjk}$ in (2) if it knows $\mathbf{R}_{jjk}$ and also $\mathbf{Q}_{jk}$. In practice, these matrices are unknown to the BS a priori and need to be estimated.

### B. Spectral Efficiency: Perfect Covariance Information

During payload transmission, the received vector $\mathbf{y}_j \in \mathbb{C}^M$ at BS $j$ is linearly processed by $\mathbf{v}_{jk} \in \mathbb{C}^M$, i.e., the receive combining vector assigned by BS $j$ to its $k$th UE. If MMSE estimation is used, then the channel capacity of UE $k$ in cell $j$ can be lower bounded by the SE [2]

$$\text{SE}_{jk} = \left(1 - \frac{K}{\tau_c}\right) \mathbb{E}\left\{\log_2\left(1 + \gamma_{jk}\right)\right\} \quad \text{[bit/s/Hz]} \quad (5)$$

where the pre-log factor accounts for pilot overhead and the instantaneous SINR is

$$\gamma_{jk} = \frac{|\mathbf{v}_{jk}^{\text{H}}\hat{\mathbf{h}}_{jjk}|^2}{\mathbf{v}_{jk}^{\text{H}}\left(\sum_{l=1}^{L}\sum_{\substack{i=1\\(l,i)\neq(j,k)}}^{K}\hat{\mathbf{h}}_{jli}\hat{\mathbf{h}}_{jli}^{\text{H}} + \mathbf{Z}_j\right)\mathbf{v}_{jk}} \quad (6)$$

with $\mathbf{Z}_j = \sum_{i=1}^{K}(\mathbf{R}_{jli} - \boldsymbol{\Phi}_{jli}) + \sum_{l=1,l\neq j}^{L}\sum_{i=1}^{K}\mathbf{R}_{jli} + \frac{1}{\rho}\mathbf{I}_M$. This is the tightest capacity bound that is available for massive MIMO systems with linear receive combining. However, this bound strictly requires the use of MMSE estimation at the BS, which in turn requires perfect knowledge of the covariance matrices. In the remainder, we consider imperfect covariance matrices and need to use an alternative capacity bound.

## III. UPLINK PERFORMANCE WITH IMPERFECT CHANNEL COVARIANCE INFORMATION

In this section, we investigate how the BS can estimate the required covariance information and we evaluate the consequences of having imperfect estimates. The channel statistics are assumed fixed over the system bandwidth $B_s$ and a time interval $T_s$. The number of coherence blocks that are contained into such a time-frequency block is

$$\tau_s = \frac{B_s T_s}{B_c T_c} = \frac{B_s T_s}{\tau_c}. \quad (7)$$

To quantify $\tau_s$, let us consider a mobile scenario with $B_s = 10$ MHz and $T_s = 0.5$ s. Each coherence block is $B_c = 200$ kHz and $T_c = 1$ ms, which allows for 135 km/h mobility at a 2 GHz carrier frequency. The channel statistics are then fixed for $\tau_s = 25000$ coherence blocks. With so many blocks, one can put some extra pilots for channel covariance estimation in a fraction of them and still keep the overhead low; see Fig. 1 for an illustration. We will now investigate how such extra pilots can be used to aid the covariance estimation.

### A. Covariance Matrix Estimation

For BS $j$ to compute the MMSE estimate of $\mathbf{h}_{jjk}$, (2) shows that it has to know $\mathbf{R}_{jjk} = \mathbb{E}\{\mathbf{h}_{jjk}\mathbf{h}_{jjk}^{\text{H}}\}$ and $\mathbf{Q}_{jk} = \mathbb{E}\{\mathbf{y}_{jk}^p(\mathbf{y}_{jk}^p)^{\text{H}}\}$. Since these are covariance matrices, the classical approach is to approximate them with the corresponding sample covariance matrices. However, since these are $M \times M$ matrices (i.e., relatively large matrices), we might need to regularize the estimates [8], [9].

*1) Estimation of $\mathbf{Q}_{jk}$:* We begin with estimation of $\mathbf{Q}_{jk}$. Suppose BS $j$ has received the pilot signal $\mathbf{y}_{jk}^p$ in (3) over $N_Q \leq \tau_s$ coherence blocks. We denote the $N_Q$ observations as $\mathbf{y}_{jk}^p[1], \ldots, \mathbf{y}_{jk}^p[N_Q]$. Note that these are obtained from the pilots already used for channel estimation—no extra pilots are needed. We can then form the sample covariance matrix as

$$\hat{\mathbf{Q}}_{jk}^{(\text{sample})} = \frac{1}{N_Q}\sum_{n=1}^{N_Q}\mathbf{y}_{jk}^p[n]\left(\mathbf{y}_{jk}^p[n]\right)^{\text{H}}. \quad (8)$$

For a particular antenna index $m$, the sample variance converges almost surely (a.s.) to the true variance as $N_Q \to \infty$:

$$\frac{1}{N_Q}\sum_{n=1}^{N_Q}\left[\mathbf{y}_{jk}^p[n]\left(\mathbf{y}_{jk}^p[n]\right)^{\text{H}}\right]_{m,m} \xrightarrow{\text{a.s.}} [\mathbf{Q}_{jk}]_{m,m}. \quad (9)$$

This follows from the law of large numbers and the ergodicity of the channels. The standard deviation of the sample variance decays as $1/\sqrt{N_Q}$, thus relatively few observations are required to obtain a good variance estimate. Each element of $\hat{\mathbf{Q}}_{jk}^{(\text{sample})}$ converges to the corresponding element of $\mathbf{Q}_{jk}$ in a similar way. It is more challenging to obtain a sample covariance matrix whose eigenvalues and eigenvectors are well aligned with those of $\mathbf{Q}_{jk}$. This is because the errors in all $M^2$ elements of $\hat{\mathbf{Q}}_{jk}^{(\text{sample})}$ influence the eigenstructure. This might have a substantial impact on massive MIMO, since the MMSE estimator exploits the eigenstructure of $\mathbf{Q}_{jk}$ to obtain a better estimate. To overcome this issue, we follow the approach proposed in [8], [9] and estimate the covariance matrix as the convex combination

$$\hat{\mathbf{Q}}_{jk}(\eta) = \eta\hat{\mathbf{Q}}_{jk}^{(\text{sample})} + (1-\eta)\hat{\mathbf{Q}}_{jk}^{(\text{diagonal})} \quad (10)$$

between the conventional sample covariance matrix and the diagonalized sample covariance matrix $\hat{\mathbf{Q}}_{jk}^{(\text{diagonal})}$, which contains the main diagonal of $\hat{\mathbf{Q}}_{jk}^{(\text{sample})}$ and otherwise is zero. Note that the main diagonal of $\hat{\mathbf{Q}}_{jk}(\eta)$ is the same for every $\eta \in [0,1]$, while the off-diagonal elements are

$\eta$ times smaller than in the sample covariance matrix. This regularization makes $\hat{\mathbf{Q}}_{jk}(\eta)$ a full-rank matrix for any $\eta < 1$, even if $N_Q < M$, and it treats unreliable off-diagonal elements by underestimating their values. If $\eta = 0$, $\hat{\mathbf{Q}}_{jk}(\eta)$ is diagonal and the correlation between the channel elements is ignored.

*2) Estimation of $\mathbf{R}_{jjk}$:* A similar approach can, in principle, be taken to estimate the individual $M \times M$ covariance matrix $\mathbf{R}_{jjk}$ of the desired UE $k$ in cell $j$. The only question is how the BS should obtain "clean" observations of $\mathbf{h}_{jjk}$, without interference from other UEs. As described in [6], a specific training phase for learning $\mathbf{R}_{jjk}$ is possible, where every UE uses a set of unique orthogonal pilots. This can be implemented using a pattern of $N_R K L$ extra pilots, as illustrated in Fig. 1, whereof each UE gets $N_R$ pilots.[1] This provides BS $j$ with $N_R$ observations of $\mathbf{h}_{jjk}$ in noise, from which a sample covariance matrix can be formed. We call this approach "R direct" and note that one needs to regularize the estimate as described above to get robustness for small $N_R$.

An alternative approach, which we propose here, is to rely on a two-stage estimation procedure. Each UE is associated with $N_R$ unique orthogonal pilots, as above, but the UE does not send the pilot itself. Instead, all other UEs that cause pilot contamination to it will send the pilot. This will allow BS $j$ to estimate the sample covariance matrix $\hat{\mathbf{Q}}_{jk,-k}^{(\text{sample})}$ of $\mathbf{Q}_{jk} - \mathbf{h}_{jjk}\mathbf{h}_{jjk}^{\mathrm{H}}$, including all pilot-contaminating interfering UEs, using the $N_R$ observations of the received pilot signals. If $\hat{\mathbf{Q}}_{jk}^{(\text{sample})}$ has been already computed, this information can be exploited to compute an estimate of the sample covariance matrix $\hat{\mathbf{R}}_{jjk}^{(\text{sample})}$ as follows:

$$\hat{\mathbf{R}}_{jjk}^{(\text{sample})} = \hat{\mathbf{Q}}_{jk}^{(\text{sample})} - \hat{\mathbf{Q}}_{jk,-k}^{(\text{sample})} \qquad (11)$$

from which the regularized estimate of $\hat{\mathbf{R}}_{jjk}$ is obtained as

$$\hat{\mathbf{R}}_{jjk}(\mu) = \mu \hat{\mathbf{R}}_{jjk}^{(\text{sample})} + (1-\mu)\hat{\mathbf{R}}_{jjk}^{(\text{diagonal})} \qquad (12)$$

where $\mu \in [0,1]$ is the regularization factor. We call this approach "Via Q". Whenever $N_Q > N_R$, which is typically the case in practice, $\hat{\mathbf{R}}_{jjk}(\mu)$ contains more observations of $\mathbf{h}_{jjk}$ than in the "R direct" approach. On the other hand, it also contains some perturbation from the imperfect subtraction of the interfering UEs' covariance matrices. We compare the two approaches numerically in the next section.

### B. Channel Estimation

By treating $\hat{\mathbf{R}}_{jjk}(\mu)$ and $\hat{\mathbf{Q}}_{jk}(\eta)$ as the true covariance matrices, we can compute an approximate MMSE estimate of $\mathbf{h}_{jjk}$ as

$$\hat{\mathbf{h}}_{jjk} = \mathbf{A}_{jjk}(\mu,\eta)\mathbf{y}_{jk}^p \qquad (13)$$

with $\mathbf{A}_{jjk}(\mu,\eta) = \hat{\mathbf{R}}_{jjk}(\mu)\big(\hat{\mathbf{Q}}_{jk}(\eta)\big)^{-1}$. Assuming that $\mathbf{y}_{jk}^p$ and $\mathbf{A}_{jjk}(\mu,\eta)$ are independent (i.e., $\mathbf{y}_{jk}^p$ is not used to estimate any of the covariance matrices), the corresponding MSE can be computed as follows.

---
[1]Each cell cannot have unique pilots in practice, but the extra pilots used for covariance estimation can be reused very sparsely in the network.

**Lemma 1.** *Consider the estimator $\hat{\mathbf{h}}_{jjk} = \mathbf{W}_{jk}\mathbf{y}_{jk}^p$, where $\mathbf{W}_{jk}$ is a deterministic matrix, then the MSE is*

$$\mathbb{E}\{\|\mathbf{h}_{jjk} - \mathbf{W}_{jk}\mathbf{y}_{jk}^p\|^2\}$$
$$= \operatorname{tr}\left((\mathbf{I}_M - \mathbf{W}_{jk} - \mathbf{W}_{jk}^{\mathrm{H}})\mathbf{R}_{jjk}\right) + \operatorname{tr}\left(\mathbf{W}_{jk}\mathbf{Q}_{jk}\mathbf{W}_{jk}^{\mathrm{H}}\right).$$

*Proof:* Follows from direct computation of the MSE. ∎

We use this lemma with $\mathbf{W}_{jk} = \mathbf{A}_{jjk}(\mu,\eta)$. The regularization factors $\mu, \eta$ can be selected to minimize the MSE.

### C. Spectral Efficiency

To quantify the SE under imperfect covariance information, we need a capacity lower bound that does not require MMSE estimates. The use-and-then-forget bound can be applied [12], where the received combined signal in (14) is rewritten as

$$\mathbf{v}_{jk}^{\mathrm{H}}\mathbf{y}_j = \mathbb{E}\{\mathbf{v}_{jk}^{\mathrm{H}}\mathbf{h}_{jjk}\}s_{jk} + \Big(\mathbf{v}_{jk}^{\mathrm{H}}\mathbf{h}_{jjk} - \mathbb{E}\{\mathbf{v}_{jk}^{\mathrm{H}}\mathbf{h}_{jjk}\}\Big)s_{jk}$$
$$+ \sum_{k=1,k\neq j}^{K}\mathbf{v}_{jk}^{\mathrm{H}}\mathbf{h}_{jjk}s_{jk} + \sum_{l=1,l\neq j}^{L}\sum_{i=1}^{K}\mathbf{v}_{jk}^{\mathrm{H}}\mathbf{h}_{jli}s_{li} + \frac{1}{\sqrt{\rho}}\mathbf{v}_{jk}^{\mathrm{H}}\mathbf{n}_j. \quad (14)$$

By treating $\mathbb{E}\{\mathbf{v}_{jk}^{\mathrm{H}}\mathbf{h}_{jjk}\}$ as a known deterministic channel and noting that the other terms are uncorrelated with $\mathbb{E}\{\mathbf{v}_{jk}^{\mathrm{H}}\mathbf{h}_{jjk}\}s_{jk}$, the following result is obtained [12].

**Lemma 2.** *The channel capacity of UE $k$ in cell $j$ is lower bounded by*

$$\underline{\text{SE}}_{jk} = \left(1 - \frac{K}{\tau_c} - \alpha\right)\log_2\left(1 + \underline{\gamma}_{jk}\right) \quad [\text{bit/s/Hz}] \quad (15)$$

*with $\alpha = \frac{N_R K L}{\tau_s}$ accounting for the extra pilots used for covariance matrix estimation and*

$$\underline{\gamma}_{jk} = \frac{|\mathbb{E}\{\mathbf{v}_{jk}^{\mathrm{H}}\mathbf{h}_{jjk}\}|^2}{\sum_{l=1}^{L}\sum_{i=1}^{K}\mathbb{E}\left\{|\mathbf{v}_{jk}^{\mathrm{H}}\mathbf{h}_{jli}|^2\right\} - \left|\mathbb{E}\{\mathbf{v}_{jk}^{\mathrm{H}}\mathbf{h}_{jjk}\}\right|^2 + \frac{1}{\rho}\mathbb{E}\{\|\mathbf{v}_{jk}\|^2\}} \quad (16)$$

*where the expectations are with respect to channel realizations.*

The lower bound in Lemma 2 is intuitively less tight than (5), since the channel estimates are only utilized for receive combining but not for signal detection. This is what the use-and-then-forget terminology refers to. The benefit of Lemma 2 is that the capacity bound does not require the use of MMSE channel estimation, but can be applied along with any channel estimator and any combining scheme. In particular, RZF with

$$\mathbf{v}_{jk} = \left(\sum_{i=1}^{K}\hat{\mathbf{h}}_{jji}\hat{\mathbf{h}}_{jji}^{\mathrm{H}} + \frac{1}{\rho}\mathbf{I}_M\right)^{-1}\hat{\mathbf{h}}_{jjk} \qquad (17)$$

is a good choice in practice [13]. Each of the expectations in (16) can be computed by Monte-Carlo simulations for any set of channel distributions. Closed-form expressions can be obtained with MRC type of schemes, expressed as $\mathbf{v}_{jk} = \mathbf{W}_{jk}\mathbf{y}_{jk}^p$. By selecting $\mathbf{W}_{jk} \in \mathbb{C}^{M \times M}$ in different ways, $\mathbf{v}_{jk}$ becomes an estimate of $\mathbf{h}_{jjk}$ with different characteristics:

$$\mathbf{W}_{jk} = \begin{cases} \mathbf{R}_{jjk}\mathbf{Q}_{jk}^{-1} & \text{MMSE estimator} \\ \mathbf{A}_{jjk}(\mu,\eta) & \text{Approximate MMSE estimator} \\ \mathbf{I}_M & \text{LS estimator.} \end{cases} \quad (18)$$

$$\underline{\gamma}_{jk} = \frac{|\text{tr}(\mathbf{W}_{jk}^{\text{H}} \mathbf{R}_{jjk})|^2}{\sum_{l=1}^{L} \sum_{i=1}^{K} \text{tr}\left(\mathbf{W}_{jk} \mathbf{Q}_{jk} \mathbf{W}_{jk}^{\text{H}} \mathbf{R}_{jli}\right) + \sum_{l=1, l \neq j}^{L} |\text{tr}(\mathbf{W}_{jk}^{\text{H}} \mathbf{R}_{jlk})|^2 + \frac{1}{\rho} \text{tr}(\mathbf{W}_{jk} \mathbf{Q}_{jk} \mathbf{W}_{jk}^{\text{H}})} \quad (24)$$

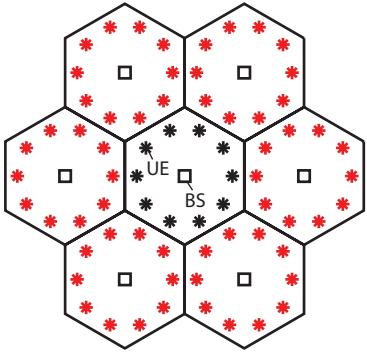

Fig. 2. Multi-cell setup with $L = 7$ cells and $K = 10$ UE in each cell, with each pilot being reused by the UE at the corresponding position in every cell.

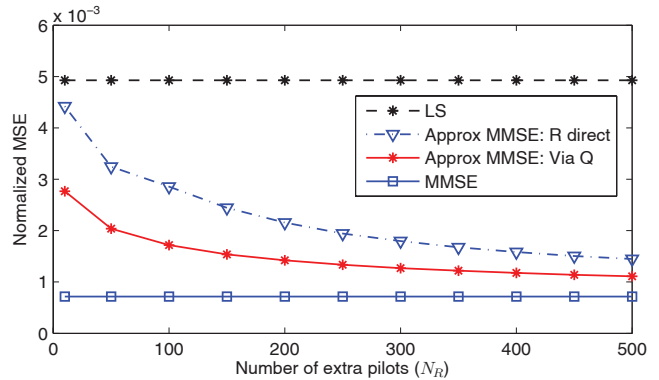

Fig. 3. Normalized MSE for intra-cell UE channels, achieved with different channel estimators for $M = 100$ in the scenario depicted in Fig. 2.

The following lemma computes the expectations in (16) for any deterministic choice of $\mathbf{W}_{jk}$. Note that $\mathbf{W}_{jk}$ is a deterministic matrix for the MMSE and LS estimators, while it can be treated as deterministic in the case of an approximate MMSE estimator, if $\mathbf{A}_{jjk}(\mu, \eta)$ is based on a different set of channel realizations than those used for data transmission (with respect to which the expectations are computed).

**Lemma 3.** *If the combining vector $\mathbf{v}_{jk} = \mathbf{W}_{jk} \mathbf{y}_{jk}^p$ is used, for some deterministic matrix $\mathbf{W}_{jk}$, then*

$$\mathbb{E}\{\mathbf{v}_{jk}^{\text{H}} \mathbf{h}_{jjk}\} = \text{tr}(\mathbf{W}_{jk}^{\text{H}} \mathbf{R}_{jjk}) \quad (19)$$
$$\mathbb{E}\{\mathbf{v}_{jk}^{\text{H}} \mathbf{v}_{jk}\} = \text{tr}(\mathbf{W}_{jk} \mathbf{Q}_{jk} \mathbf{W}_{jk}^{\text{H}}) \quad (20)$$

*and*

$$\mathbb{E}\left\{|\mathbf{v}_{jk}^{\text{H}} \mathbf{h}_{jli}|^2\right\} = \text{tr}\left(\mathbf{W}_{jk} \mathbf{Q}_{jk} \mathbf{W}_{jk}^{\text{H}} \mathbf{R}_{jli}\right) + \begin{cases} 0 & \text{if } i \neq k \\ \left|\text{tr}(\mathbf{W}_{jk}^{\text{H}} \mathbf{R}_{jli})\right|^2 & \text{if } i = k. \end{cases} \quad (21)$$

*Proof:* The proof is available in the appendix. ∎

By utilizing these results, we obtain a closed-form expression for the SE with MRC using an arbitrary channel estimator.

**Theorem 1.** *If the MRC vector $\mathbf{v}_{jk} = \mathbf{W}_{jk} \mathbf{y}_{jk}^p$ is used, for some deterministic matrix $\mathbf{W}_{jk}$, then the channel capacity of UE $k$ in cell $j$ is lower bounded by* (15) *with $\underline{\gamma}_{jk}$ given in* (24) *at the top of the page.*

## IV. NUMERICAL EVALUATION

We consider the urban multi-cell scenario in Fig. 2, with $L = 7$ cells and 300 m inter-BS distance. The uplink performance is evaluated in the center cell where the BS has $M = 100$ antennas. Each cell contains $K = 10$ UEs at 120 m distance from the serving BS, in a circular pattern, and the UEs that are at the same position in different cells reuse the same pilot. Inspired by the 3GPP pathloss models, the SNR at distance $d$ m from a BS is computed as $78.7 - 37.6 \log_{10}(d)$ dB. The channel coherence is modeled as exemplified in the beginning of Section III: each coherence block has $B_c = 200$ kHz and $T_c = 1$ ms such that $\tau_c = 200$. With $B_s = 10$ MHz and $T_s = 0.5$ s, we have $\tau_s = 25000$.

We will illustrate the estimation performance by using the one-ring model for the channel covariance matrices. This model describes a uniform linear array (here with half-wavelength antenna separation) where the multipath components from a UE arrive uniformly distributed in an angular interval (here 20°) centered around the geographical angle to the UE. This model has been used in the massive MIMO literature to show the impact of strong spatial correlation [5], [6], but we stress that results similar to those presented in this section can be obtained with other covariance models as well.

The normalized MSE, $\mathbb{E}\{\|\mathbf{h}_{jjk} - \mathbf{W}_{jk} \mathbf{y}_{jk}^p\|^2\}/\text{tr}(\mathbf{R}_{jjk})$, is shown in Fig. 3 for an average UE in the center cell. The horizontal axis shows different numbers of extra pilots $N_R$ used to estimate $\mathbf{R}_{jjk}$. For every such value, we assume that $N_Q = 10 N_R$, since every coherence block can be used to estimate $\mathbf{Q}_{jk}$. The approximate MMSE estimator (with optimized $\eta$ and $\mu$) is implemented using either the method in [6], which estimates $\mathbf{R}_{jjk}$ directly, or the proposed approach, which operates as in (11)–(12). Comparisons are made with the LS and MMSE estimators. We notice that $N_R = 10$ is sufficient for both approaches to outperform the LS estimator. The MSE reduces monotonically with $N_R$ and a few hundred samples are required to achieve performance close to the MMSE estimator. The proposed estimation approach "Via Q" largely outperforms "R direct" from [6] for all values of $N_R$. The main reason is that the estimate of $\mathbf{R}_{jjk}$ contains $N_Q$ realizations of $\mathbf{h}_{jjk}$, instead of $N_R$ as in "R direct".

Next, the impact that imperfect channel covariance information has on the SE in the center cell is evaluated, using either MRC or RZF. Fig. 4 shows the sum SE as a function of $N_R$. Firstly, we notice that the LS estimator performs reasonably well, without knowing the channel covariance matrices. In particular, 82% of the SE obtained with the MMSE estimator is achieved for MRC, while 70% of the SE is achieved for RZF. Secondly, the proposed approximate MMSE estimator "Via Q" outperforms the LS estimator when using as little as $N_R = 10$ extra pilots. To achieve 95% of the SE of the MMSE estimator, $N_R = 100$ is needed with MRC and $N_R = 250$ with RZF. The gain in estimation quality clearly outweighs the extra pilot overhead (larger $\alpha$ in (15)). Thirdly, the approach "R direct" from [6] performs poorly when $N_R$ is small (roughly around a few tens of extra pilots). When used with RZF, the value of $N_R$ needed to achieve 95% of the SE of the MMSE estimator is increased by a factor 2.5 compared to the proposed scheme. The performance gap is somewhat smaller when using MRC.

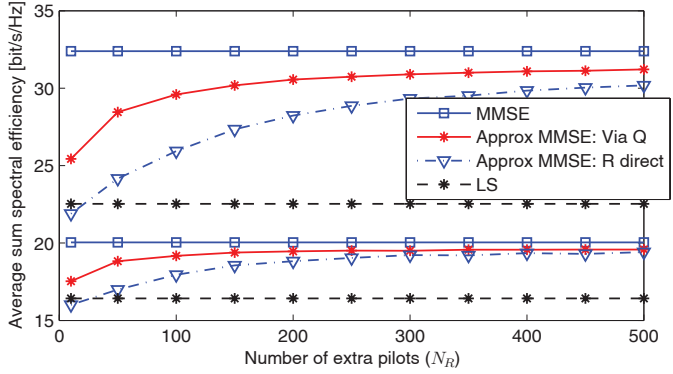

Fig. 4. Sum SE achieved with MRC and RZF, with different channel estimators. The performance is evaluated in the center cell of Fig. 2.

## V. Conclusions

We discussed how a BS can estimate the covariance information needed in massive MIMO systems for MMSE channel estimation. We proposed a new covariance estimation approach, which outperforms previous solutions under the same pilot overhead. We derived MSE and SE expressions that apply under imperfect covariance information, for any combining scheme. A closed-form SE expression was computed for MRC. Numerical results showed that covariance knowledge is not critical, but largely improves the system performance, especially when interference suppression schemes, such as RZF, are used. Only a few tens of extra pilots are sufficient to outperform a system using the LS estimator (for which no covariance information is needed) either with MRC or RZF.

## Appendix

**Proof of Lemma 3:** The expectations in (19)–(21) are computed using channel statistics. First, we prove (19) as

$$\mathbb{E}\{\mathbf{v}_{jk}^{\text{H}}\mathbf{h}_{jjk}\} = \mathbb{E}\{(\mathbf{y}_{jk}^p)^{\text{H}}\mathbf{W}_{jk}^{\text{H}}\mathbf{h}_{jjk}\} = \mathbb{E}\{\mathbf{h}_{jjk}^{\text{H}}\mathbf{W}_{jk}^{\text{H}}\mathbf{h}_{jjk}\}$$
$$= \text{tr}(\mathbf{W}_{jk}^{\text{H}}\mathbb{E}\{\mathbf{h}_{jjk}\mathbf{h}_{jjk}^{\text{H}}\}) = \text{tr}(\mathbf{W}_{jk}^{\text{H}}\mathbf{R}_{jjk}) \quad (25)$$

where the second equality follows from the fact that $\mathbf{h}_{jjk}$ is independent of the noise and all other channels and the third equality utilizes the following rule: $\mathbf{x}^{\text{H}}\mathbf{y} = \text{tr}(\mathbf{y}\mathbf{x}^{\text{H}})$ for any vectors $\mathbf{x}, \mathbf{y}$. Next, we prove (20) by noting that

$$\mathbb{E}\{\mathbf{v}_{jk}^{\text{H}}\mathbf{v}_{jk}\} = \mathbb{E}\{(\mathbf{y}_{jk}^p)^{\text{H}}\mathbf{W}_{jk}^{\text{H}}\mathbf{W}_{jk}\mathbf{y}_{jk}^p\} =$$
$$= \text{tr}(\mathbf{W}_{jk}\mathbb{E}\{\mathbf{y}_{jk}^p(\mathbf{y}_{jk}^p)^{\text{H}}\}\mathbf{W}_{jk}^{\text{H}}) = \text{tr}(\mathbf{W}_{jk}\mathbf{Q}_{jk}\mathbf{W}_{jk}^{\text{H}}) \quad (26)$$

by using the same trace rule as above and identifying $\mathbf{Q}_{jk}$ from (4). Finally, we need to compute (21) and begin with $i \neq k$, which implies that $\mathbf{v}_{jk}$ and $\mathbf{h}_{jli}$ are independent. Direct computation yields

$$\mathbb{E}\{|\mathbf{v}_{jk}^{\text{H}}\mathbf{h}_{jli}|^2\} = \mathbb{E}\left\{\mathbf{v}_{jk}^{\text{H}}\mathbf{R}_{jli}\mathbf{v}_{jk}\right\}$$
$$= \text{tr}(\mathbf{W}_{jk}\mathbb{E}\{\mathbf{y}_{jk}^p(\mathbf{y}_{jk}^p)^{\text{H}}\}\mathbf{W}_{jk}^{\text{H}}\mathbf{R}_{jli}) = \text{tr}\left(\mathbf{W}_{jk}\mathbf{Q}_{jk}\mathbf{W}_{jk}^{\text{H}}\mathbf{R}_{jli}\right) \quad (27)$$

by exploiting the independence and then following the same procedure as in (26). If $i = k$, $\mathbf{v}_{jk}$ and $\mathbf{h}_{jli}$ are dependent, but we notice that $\mathbf{v}_{jk} - \mathbf{W}_{jk}\mathbf{h}_{jli}$ and $\mathbf{h}_{jli}$ are independent. We utilize this to compute $\mathbb{E}\{|\mathbf{v}_{jk}^{\text{H}}\mathbf{h}_{jli}|^2\}$ as

$$\mathbb{E}\left\{|(\mathbf{v}_{jk} - \mathbf{W}_{jk}\mathbf{h}_{jli})^{\text{H}}\mathbf{h}_{jli}|^2\right\} + \mathbb{E}\left\{|\mathbf{h}_{jli}^{\text{H}}\mathbf{W}_{jk}\mathbf{h}_{jli}|^2\right\}$$
$$= \text{tr}\left(\mathbf{W}_{jk}(\mathbf{Q}_{jk} - \mathbf{R}_{jli})\mathbf{W}_{jk}^{\text{H}}\mathbf{R}_{jli}\right)$$
$$+ |\text{tr}(\mathbf{W}_{jk}^{\text{H}}\mathbf{R}_{jli})|^2 + \text{tr}\left(\mathbf{W}_{jk}\mathbf{R}_{jli}\mathbf{W}_{jk}^{\text{H}}\mathbf{R}_{jli}\right)$$
$$= \text{tr}\left(\mathbf{W}_{jk}\mathbf{Q}_{jk}\mathbf{W}_{jk}^{\text{H}}\mathbf{R}_{jli}\right) + |\text{tr}(\mathbf{W}_{jk}^{\text{H}}\mathbf{R}_{jli})|^2 \quad (28)$$

where the first equality utilizes the independence and the second one utilizes [14, Lemma 3] and the procedure in (27).